# ATLAS Monitored Drift Tube Chambers in E = 11 MeV Neutron Background

T. Müller, A. Mlynek, O. Biebel, R. Hertenberger, T. Nunnemann, D. Merkl, F. Rauscher, D. Schaile, R. Ströhmer

***Abstract*—The influence of fast neutrons on the occupancy and the single tube resolution of ATLAS muon drift detectors was investigated by exposing a chamber built out of 3 layers of 3 short standard drift tubes to neutron flux-densities of up to 16 kHz/cm$^2$ at a neutron energy of E=11 MeV. Pulse shape capable NE213 scintillaton detectors and a calibrated BF3 neutron detector provided monitoring of the neutron flux-density and energy. The sensitivity of the drift chamber to the neutrons was measured to be 4*10$^{-4}$ by comparing data sets with and without neutron background. For the investigation of tracks of cosmic muons two silicon-strip detectors above and underneath the chamber allow to compare measured drift-radii with reference tracks. Alternatively, the single tube resolution was determined using the triple-sum method. The comparison between data with and without neutron irradiation shows only a marginal effect on the resolution and little influence on the muon track reconstruction.**

## I. Introduction

Monitored drift tube (MDT) chambers [1] are used for high precision tracking in the ATLAS muon spectrometer. A position resolution of about 80 $\mu$m is required for momentum resolution of $\frac{\Delta p}{p} \leq 10\%$ at 1 TeV muon momentum. Neutrons and photons from beam related interactions will contribute a substantial background and will worsen the single tube resolution. Fig. 1 shows for the barrel part of ATLAS simulated flux densities between 2 and 5 $\frac{\text{kHz}}{\text{cm}^2}$ at nominal luminosity of $10^{34}$ / (cm$^2$ s) [2]. After luminosity upgrade by a factor of 10 from LHC to superLHC the track reconstruction might fail in parts of the forward spectrometer completely due to occupancy problems. The influence of $\gamma$ irradiation has been studied before [3]. Here, in the dedicated setup shown in Fig. 2, the sensitivity of MDT chambers to neutron radiation [4] and the impact of neutron radiation on the resolution of monitored drift tube (MDT) chambers [5] was investigated. At ATLAS the neutron energy spectrum covers many orders of magnitudes. In this study we concentrated on the so far unexplored region of fast neutrons above 1 MeV energy.

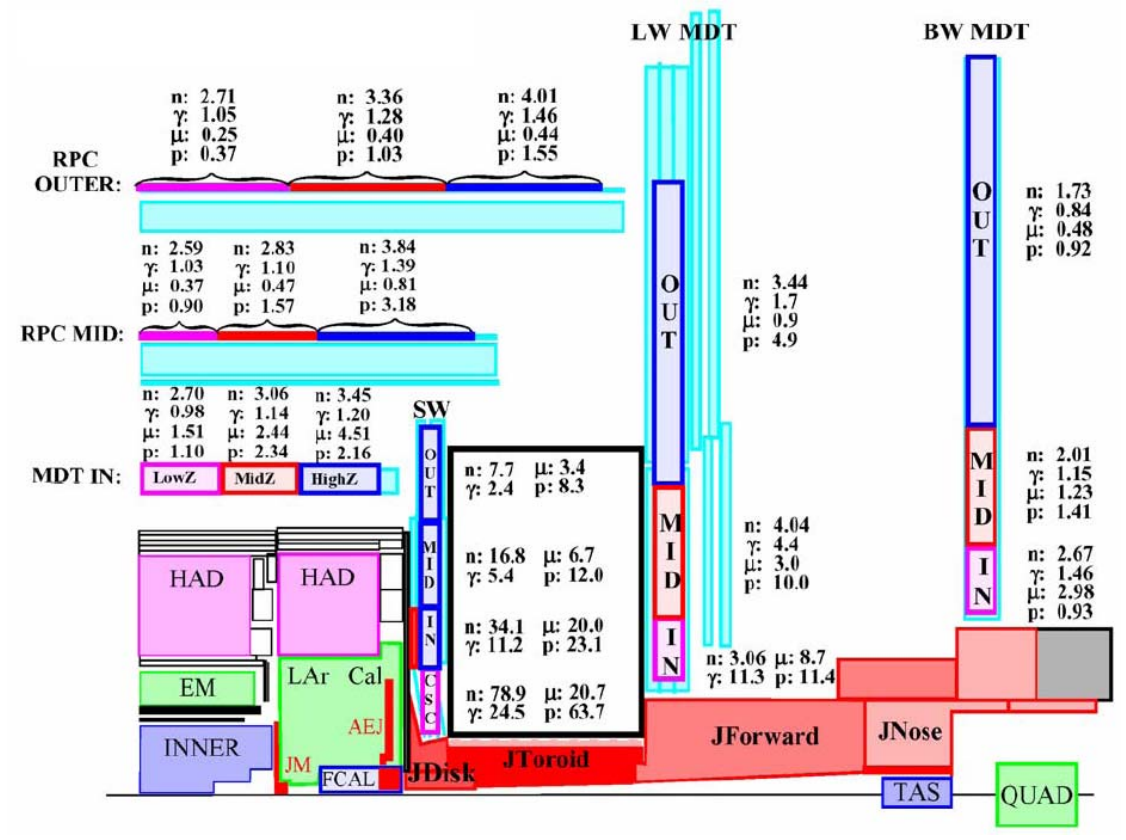

Fig. 1. Estimate of the background flux-densities in the ATLAS muon spectrometer from [2] for the nominal luminosity of $10^{34}$ / (cm$^2$ s). The units for n and $\gamma$ are [kHz / cm$^2$ s] for p and $\mu$ [Hz / cm$^2$ s].

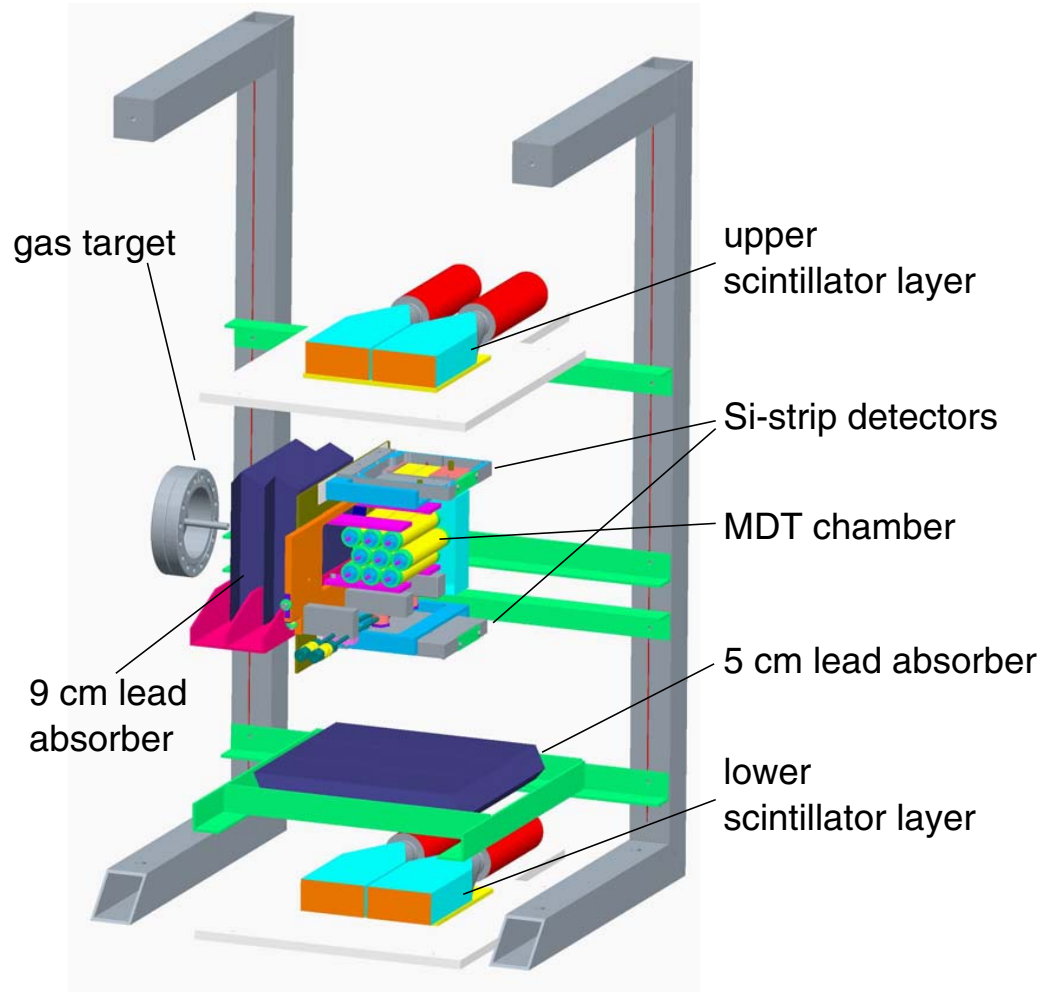

Fig. 2. Setup to study a small MDT chamber under neutron irradiation.

## II. Neutron detection probability

High energy neutrons were produced using the $^{1}$H($^{11}$B,n)$^{11}$C reaction at the tandem accelerator of the Munich Maier-Leibnitz laboratory. The 60 MeV $^{11}$B beam yields nearly mono-energetic neutrons of 11 MeV into the acceptance cone of about 30° using as target a hydrogen filled stainless steel cylinder, 3 cm long and 2 cm in diameter at 3 bar absolute pressure. A 3.5 $\mu$m thick molybdenum entrance foil was selected for low background neutron production. The noninteracting $^{11}$B beam was dumped in a 2 mm thick golden beamstop. The complete target was electrically insulated against the beam line to allow monitoring of the beam current using a current meter. Gamma irradiation created in the target cell or in the beamstop was absorbed in 9 cm of lead shielding between the $H_2$ target cell and the MDT chamber. This eliminated nearly all gamma background. The

LMU München, Fakultät für Physik, Am Coulombwall 1, D-85748 Garching, Germany, R.Hertenberger@physik.uni-muenchen.de

MDT chamber with 9 tubes of 15 cm length was equipped with the same kind of drift tubes and readout electronics as used for the muon spectrometer in the ATLAS detector: a 30 mm diameter aluminum tube of 0.4 mm wall thickness houses a gold plated W-Re anode wire of 50 $\mu$m diameter. Also the same operating conditions were chosen: 3080 V for the drift field and a gas mixture of 93% Ar and 7% $CO_2$ at 3 bar absolute pressure. The chamber was placed about 15 cm behind the $H_2$ target. This defines the acceptance angle of about 30° for the 11 MeV neutrons.

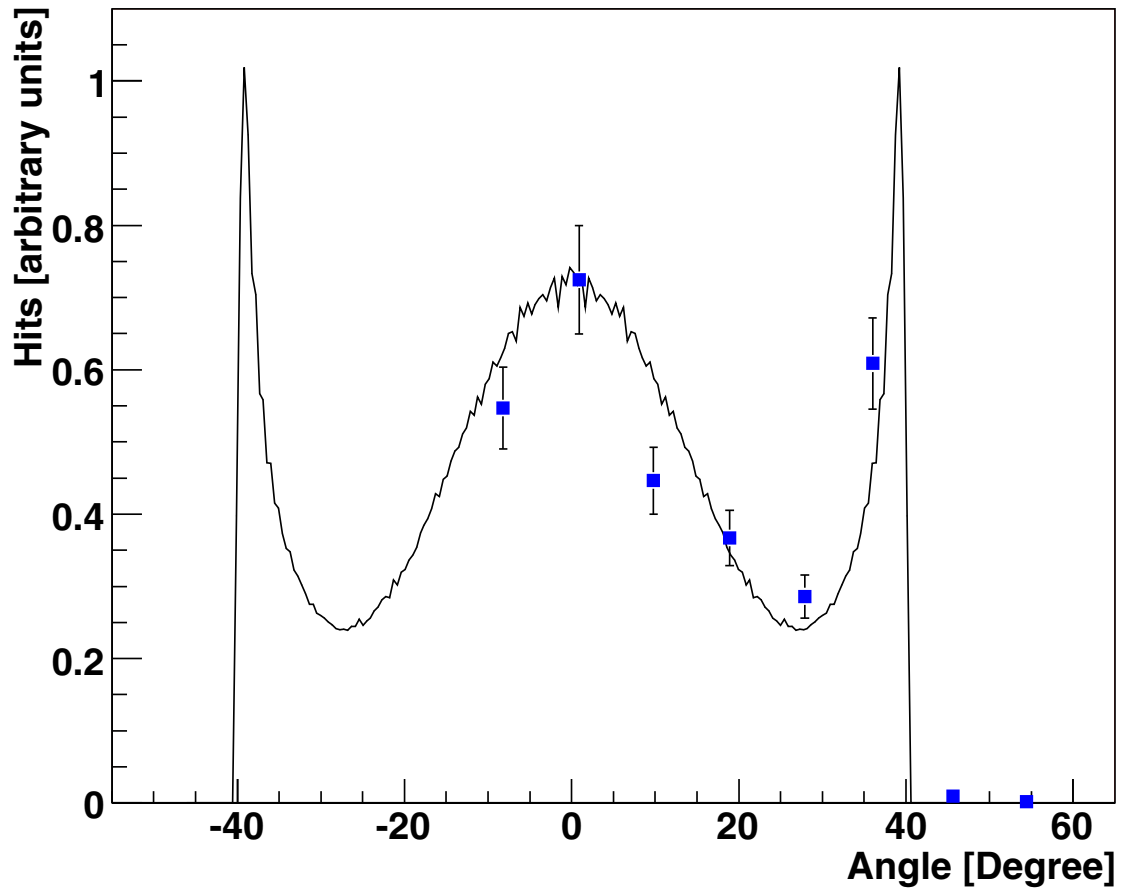

Fig. 3. Comparison of simulated and measured angular distributions of the produced neutrons.

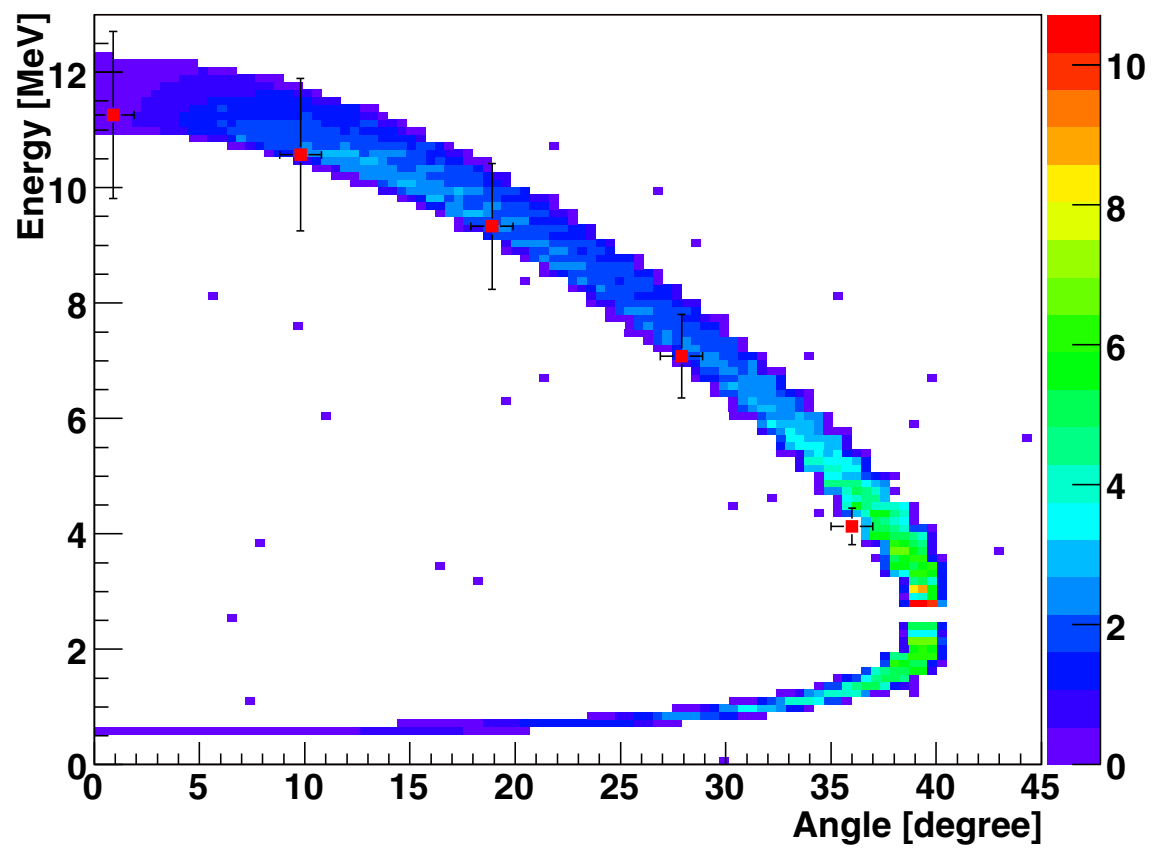

Fig. 4. Comparison of simulated and measured energy-angle distributions of the produced neutrons.

To measure the neutron energy by time of flight and the angular distribution of the neutrons four NE213 liquid scintillators were used at pulsed ion beam. Those counters allow to distinguish gamma from neutron induced signals using pulse-shape-discrimination (PSD). The absolute neutron flux was monitored with a calibrated $BF_3$ neutron radiation counter. The influence of the lead absorber in front of the MDT chambers on the neutron flux through each of the nine tubes was simulated with Geant4 [6] using as input the differential cross sections of the $^{1}H(^{11}B,n)^{11}C$ reaction from the code DROSG2000[7]. The results are compared with the measurement from the NE213 counters. Due to scattering in the lead the useable neutron flux is reduced by 56%. Fig. 3 and 4 show the agreement between simulation and measurement. The total flux-density of in this case 5.6 $\frac{\mathrm{kHz}}{\mathrm{cm}^2}$ agrees within the errors with the measurement of the $BF_3$ counter placed 2.6 m away from the target cell.

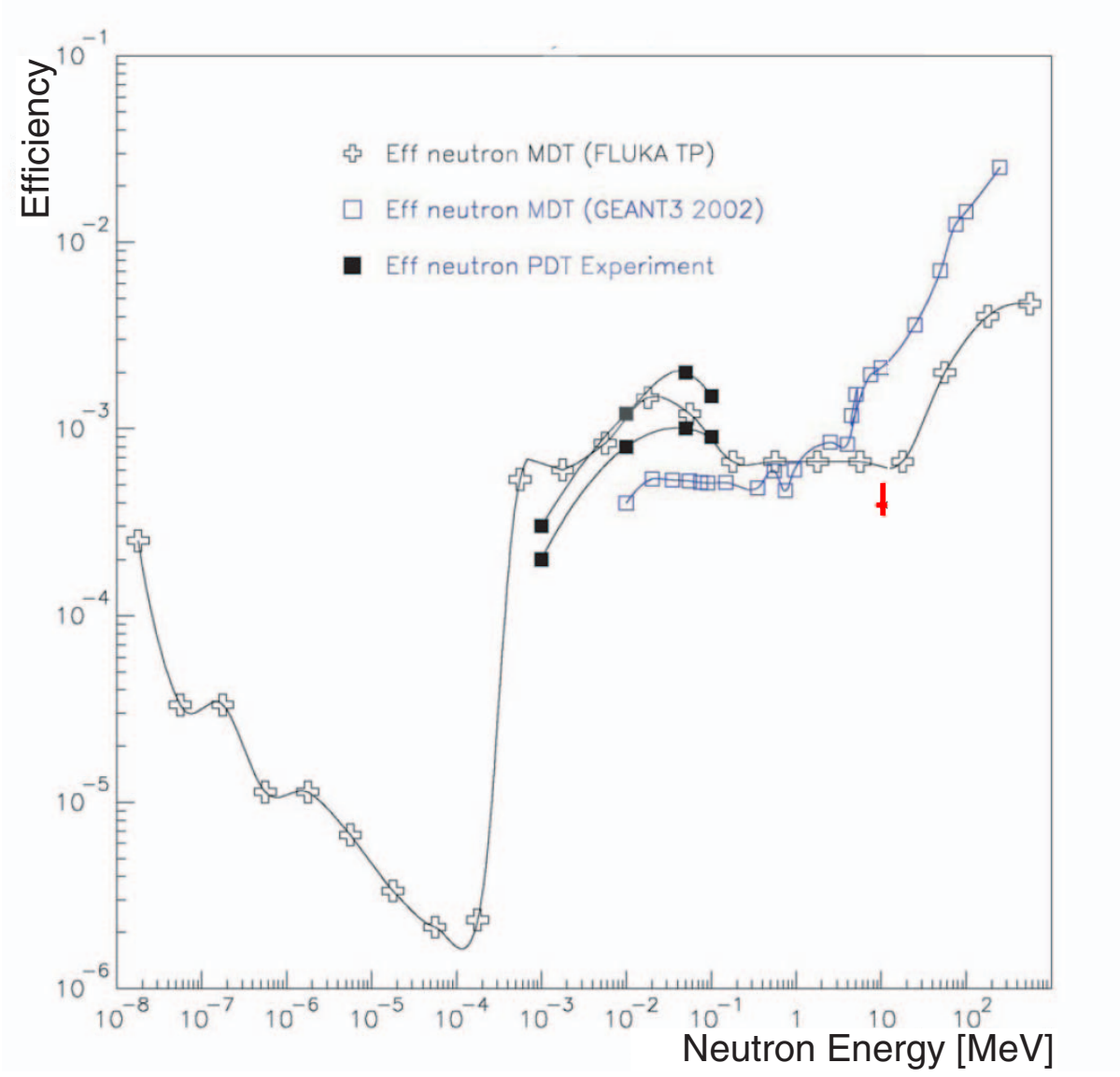

Fig. 5. Different simulations of neutron efficiency. Our result is marked in red

To determine the overall neutron efficiency, that quantifies the sensitivity of the MDT chamber to neutrons, the MDT chamber readout was randomly triggered. For these and all other irradiations dc ion beam was used. The total efficiency results from the known time window of 800 ns for a random neutron interaction, the total flux of neutrons, the measured beam current and the simulations described above. In order to subtract background of neutrons produced e.g. in the entrance foil or in the beamstop we compared measurements at 3 bar of $H_2$ with those at evacuated target cell.

The efficiency of $(4.0^{+1.6}_{-0.3}) \cdot 10^{-4}$ obtained from our measurement is indicated in Fig. 5. The error includes statistical and systematic uncertainties. Two simulations [2] are shown as well. They differ by a factor of 4 in the neutron energy range around 11 MeV where no experimental measurement existed up to now. Our measurement favors the FLUKA simulation [8].

## III. Influence of 11 MeV Neutrons on the Single Tube Resolution

For the investigation of the impact of neutron radiation on the resolution of MDT chambers a five centimeter thick lead absorber was put between the 9 drift tubes and the lower trigger scintillator to suppress the cosmic muon flux with energies less than 100 MeV. Higher energy muons were triggered by a coincidence of scintillating detectors. The position of a cosmic muon was reconstructed by employing two silicon strip detectors (Fig. 2) located close to the MDT chamber. An accuracy of about 10 $\mu$m could be achieved when adjusting the 20 $\mu$m silicon strips parallel to the tubes. The position resolution of the silicon strip detectors is better than 20 $\mu$m and thus negligible in comparison to the resolution of the drift chamber. The average neutron flux in the tubes of the MDT chamber ranged in this part of the experiment from 4.4 to 16.0 kHz/cm$^2$. This flux density corresponds to about 2 to 10 times the expected flux of neutrons with energies of at least 10 MeV in the barrel part of the ATLAS muon spectrometer after a tenfold luminosity upgrade of LHC to SLHC. Some regions of the endcap part of the ATLAS muon spectrometer are expected to suffer even higher neutron fluxes.

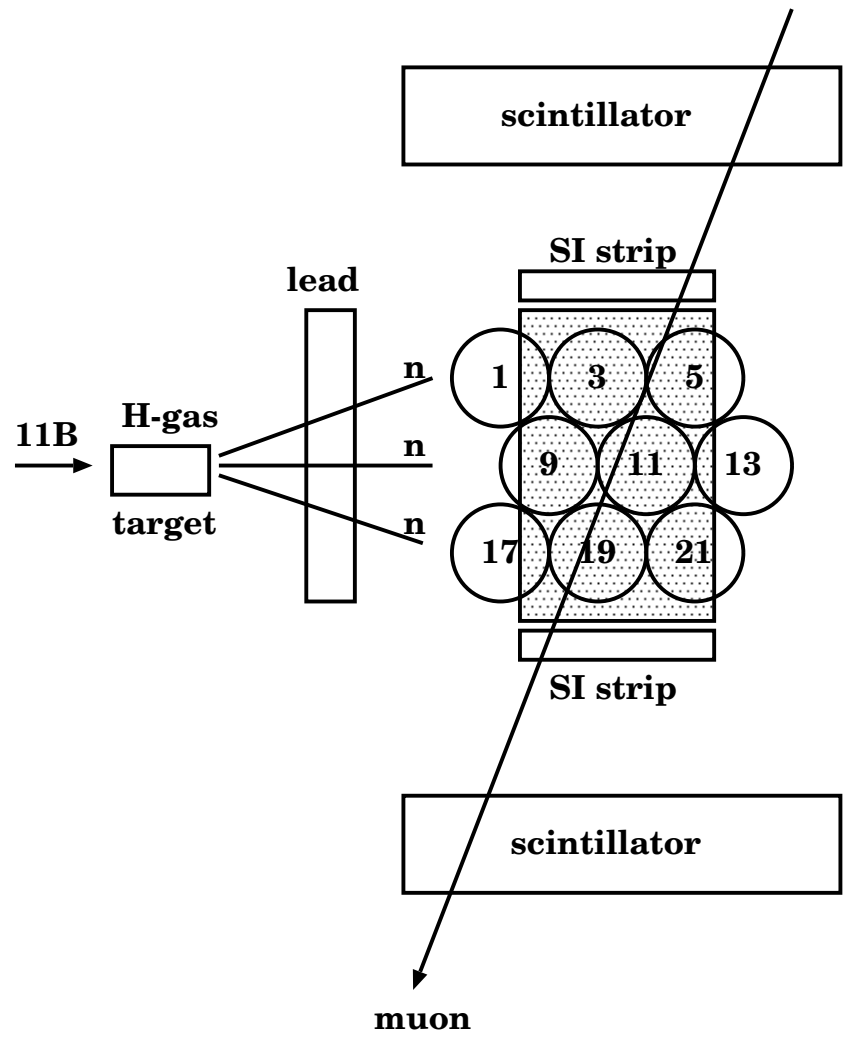

Fig. 6. Illumination of drift tubes by neutrons from the gas target. The shaded area indicates the acceptance of the silicon-strip telescope for cosmic muons.

Data on the resolution degradation could be collected over roughly two days at the tandem accelerator. Although the statistics were limited, a measurement of the degradation of spatial resolution due to the neutron irradiation was possible for five of the nine MDT tubes, see Fig. 6 together with Fig. 7. Those five tubes showed effects of up to three standard deviations. The spread in the spatial resolution degradation is partly due to the different illumination of the tubes and possibly also due to temperature variation during the measurement period.

Fig. 7 shows the drift-time - drift-radius relations (r-t relation) for all 9 tubes. The track radius as derived from the silicon detectors (x-axis) is compared to the drift-time measured by the drift tubes (y-axis). The V-shaped plots are expected from the nonlinear Ar/CO2 driftgas mixture with maximum drift-times of about 700 ns. The extended maximum drift-time of tube 5 indicates a noncentered anode wire or insufficient mechanical wire tension. For this reason this tube is not considered in the analysis. Points below the 'V' are due to random background events or delta electrons created by the muons in the active volume of the drift tubes. In both cases the measured drift-time should be below the expectation from the track. 9 fits by polynomials of eighth order to the 'V' shapes provide for each tube the transformation of measured drift-times into drift-radii. Fig. 8 shows for tube 3 the residuals, i.e. the difference between the measured drift-radii $d_i$ and the radius prediction of the silicon detectors $d_{track}$. Multiple gaussian functions are needed to describe the peak as the events from early background hits or delta-electrons smear out the distribution and produce broader tails. A double-gauss function

$$f(x) = \frac{a_0}{\sqrt{2\pi}\,w}\,e^{-\frac{1}{2}\left(\frac{x-\mu}{w}\right)^2} + \frac{a_0}{\sqrt{2\pi}\,3w}\,e^{-\frac{1}{2}\left(\frac{x-\mu}{3w}\right)^2}$$

was able to fit the residual distribution using the widths $w$ and $3w$. The width $w$ is then a measure for the tube resolution. Fig. 9 shows the resulting tube resolution as a function of drift-radius $d$. The degradation of the resolution at small radii is due to the minimum ionizing energy loss of the muons creating clusters of primary electrons in the counting gas only every 200 $\mu$m on average. This has the largest impact on the resolution when the muon passes the tube next to the wire. The resolution curve from Fig. 9 differs from the result in Ref. [3] by an offset of about 40 $\mu$m. This is due to the increased contribution of multiple scattering in the walls of the 3 tubes hit by low energy muons of momenta as low as 100 MeV [5], [9]. The multiple scattering of the 100 GeV muons used in Ref. [3] is negligible with respect to the tube resolution.

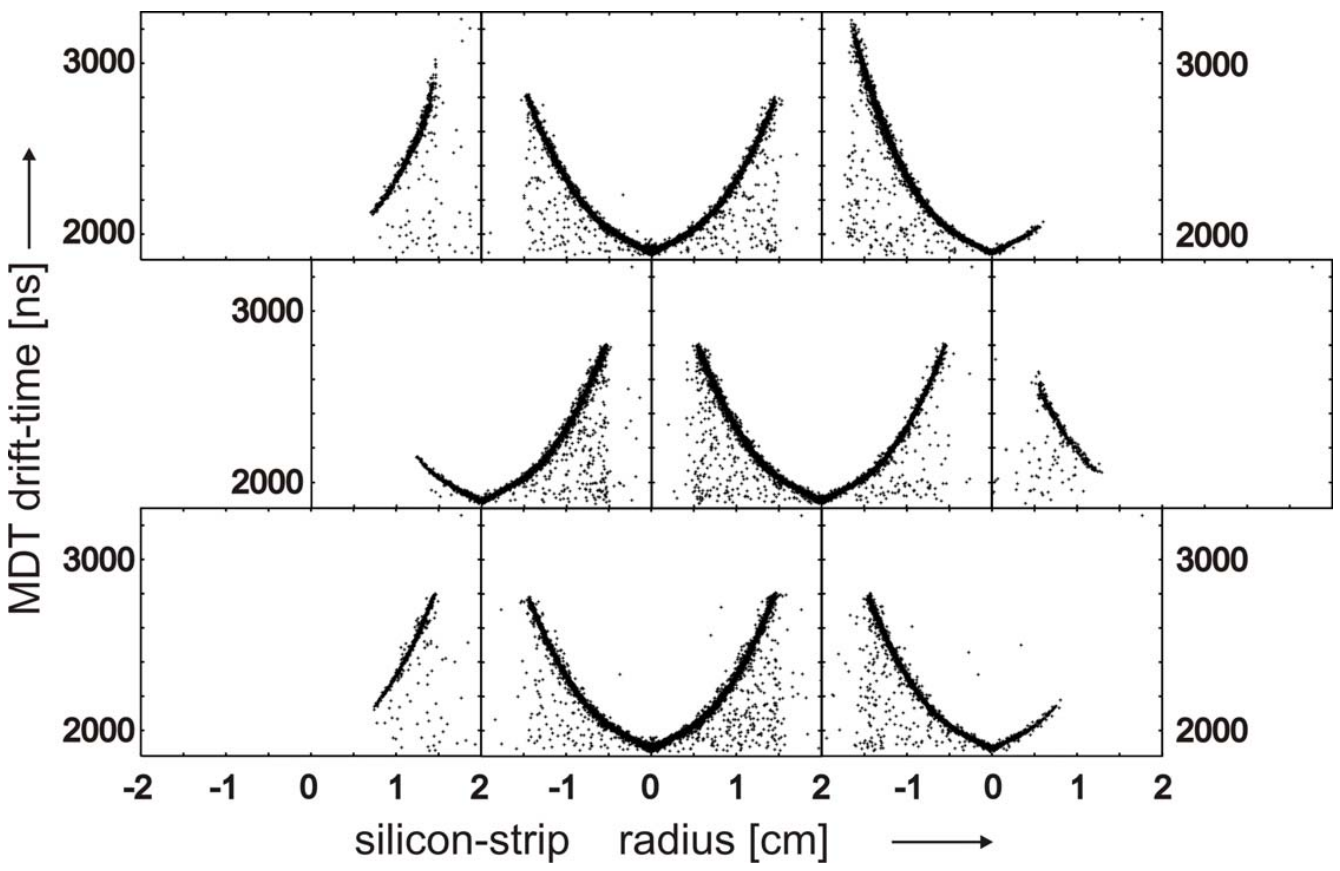

Fig. 7. Space - drift-time relation from cosmic muons for drift tubes inside the silicon strip detector acceptance.

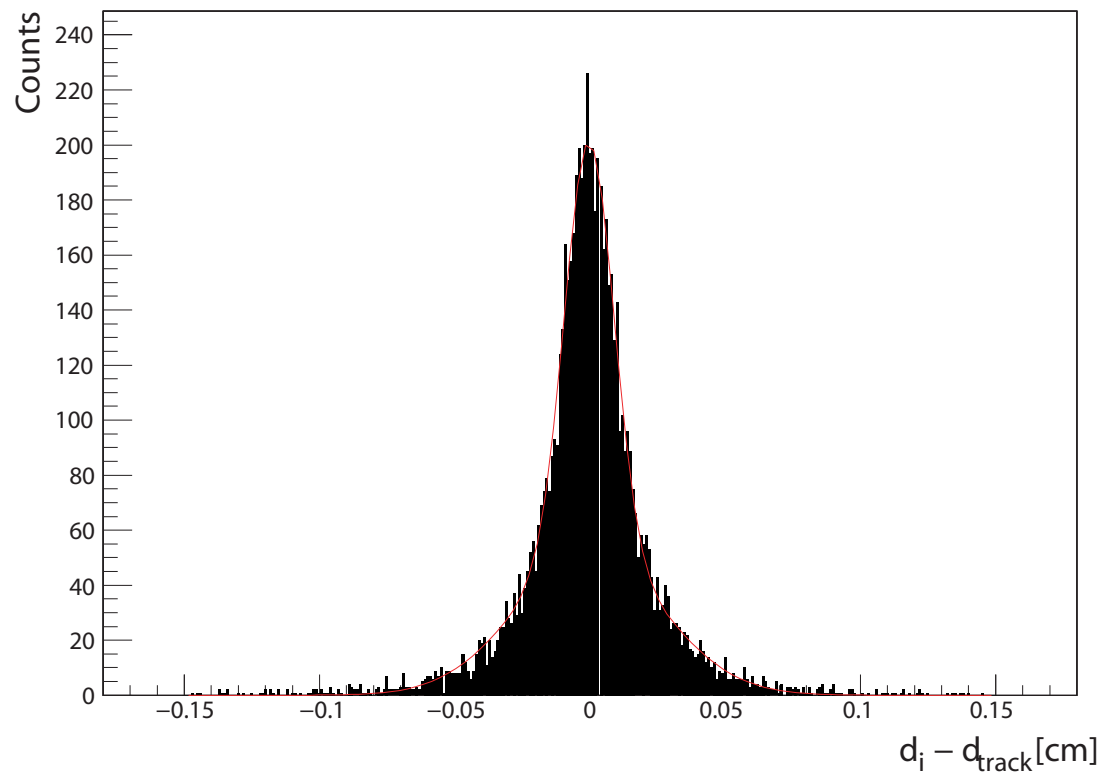

Fig. 8. Difference between the measured drift-radius and the radius predicted by the silicon detectors. A double-gauss function is used for fitting.

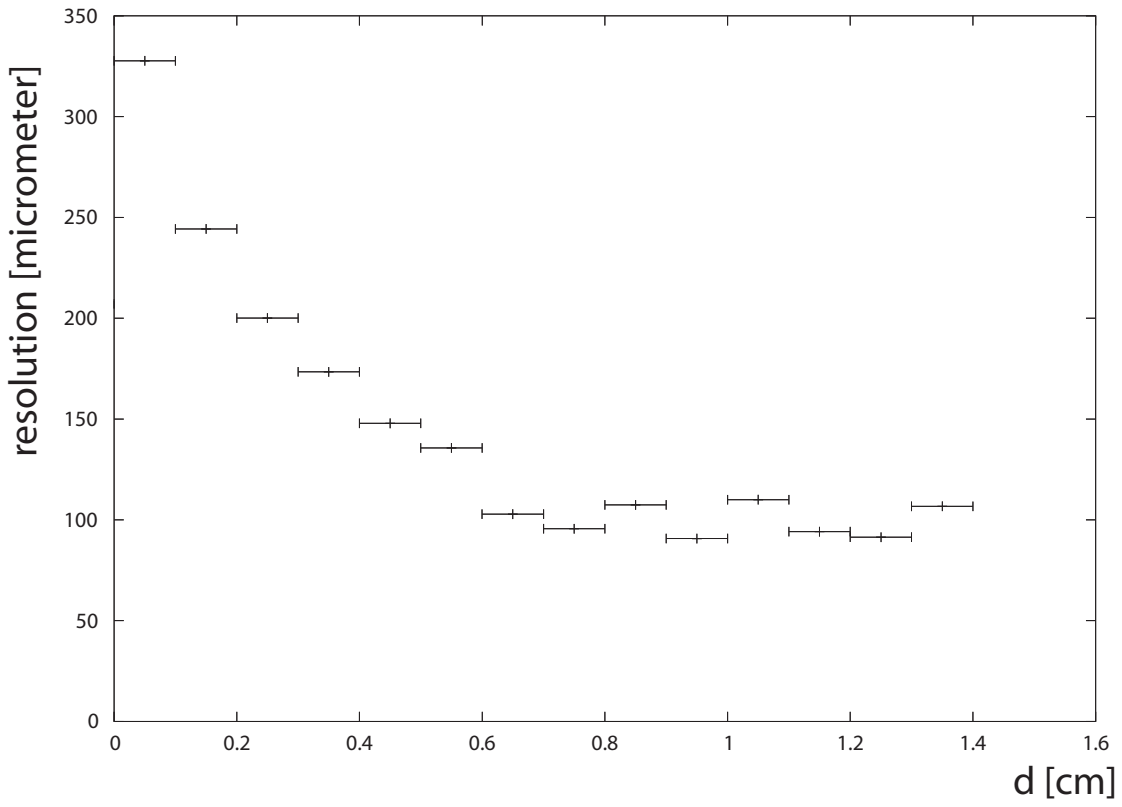

Fig. 9. Single tube resolution as a function of drift-radius without neutron background.

## IV. Triple Sum Method

Alternatively, for muons hitting a triplet of the tubes 3,9,19 or 3,11,19, see Fig. 6, the correlation between the three drift-radii and the muon track fitted to these three radii was used to determine the single tube resolution.

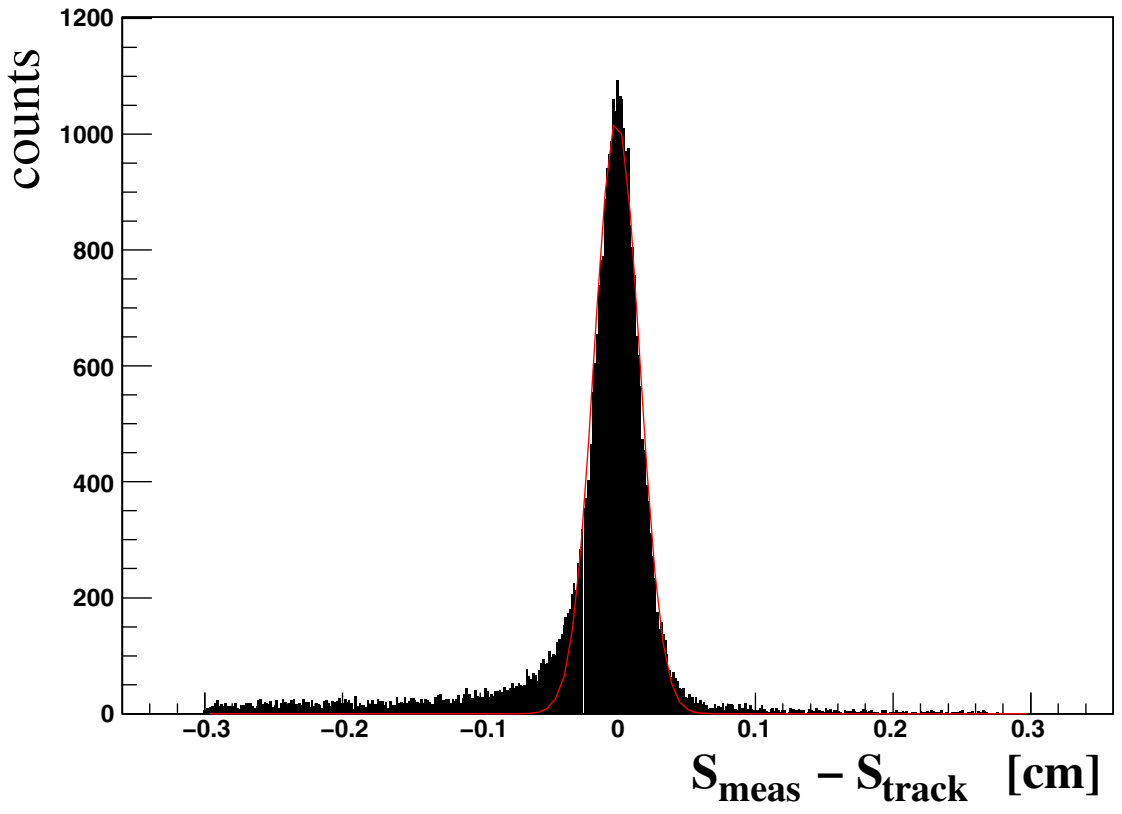

Fig. 10. Correlation of the sum of 3 measured drift-radii $S_{meas}$ with the respective theoretical expectation from the fitted track $S_{track}$. The extended tail to negative values comes from the early delta-electron or random background events.

The effective sum $S_{meas}$

$S_{meas} = \frac{1}{2}d_3 - d_{9,11} + \frac{1}{2}d_{19}$

where the sign of the measured drift-radii $d_i$ has to be taken into account, is compared to the quantity $S_{track}$ of the radii derived from the reconstructed muon track. The distance $d$ from a wire for a straight track

$x = a \cdot y + b$

where vertical tracks have slope 0, is given by:

$d = \frac{1}{\sqrt{a^2+1}}(a \cdot y + b - x)$

This defines the effective triple sum derived from the track to

$S_{track} = \frac{1}{\sqrt{a^2+1}} \left[ \left(x_2 - \frac{1}{2}(x_1 + x_3)\right) + a \left(\frac{1}{2}(y_1 + y_3) - y_2\right) \right]$

with the slope $a$ of the fitted track and the positions $x_i$, $y_i$ of the 3 wires. The difference $S_{meas} - S_{track}$ is displayed in Fig. 10. The width $\sigma_S$ of the distribution is related to the single tube resolution by

$\sigma_S = \sqrt{\frac{3}{2}}\, \sigma_{tube}$

when assuming identical tube resolution for each tube.

The tube resolution derived from the gaussian fit to Fig. 10 is shown in Fig. 11 by the triangles. As in this case at least one small drift-radius enters into the analysis, this method results in slightly increased single tube resolutions.

The full circle symbols in Fig. 11 show the single tube resolution for the five tubes 3, 9, 11, 19, and 21 derived from the comparison with the silicon strip detectors. From a statistical analysis it is concluded that the quadratic contribution due to neutron irradiation to the spatial resolution of a single tube is less than 54 $\mu$m at 95% confidence level. So, the resolution of a single tube degrades from the nominal 100 $\mu$m to about 114 $\mu$m due to neutron irratiation with fluxes of up to 16 kHz/cm$^2$.

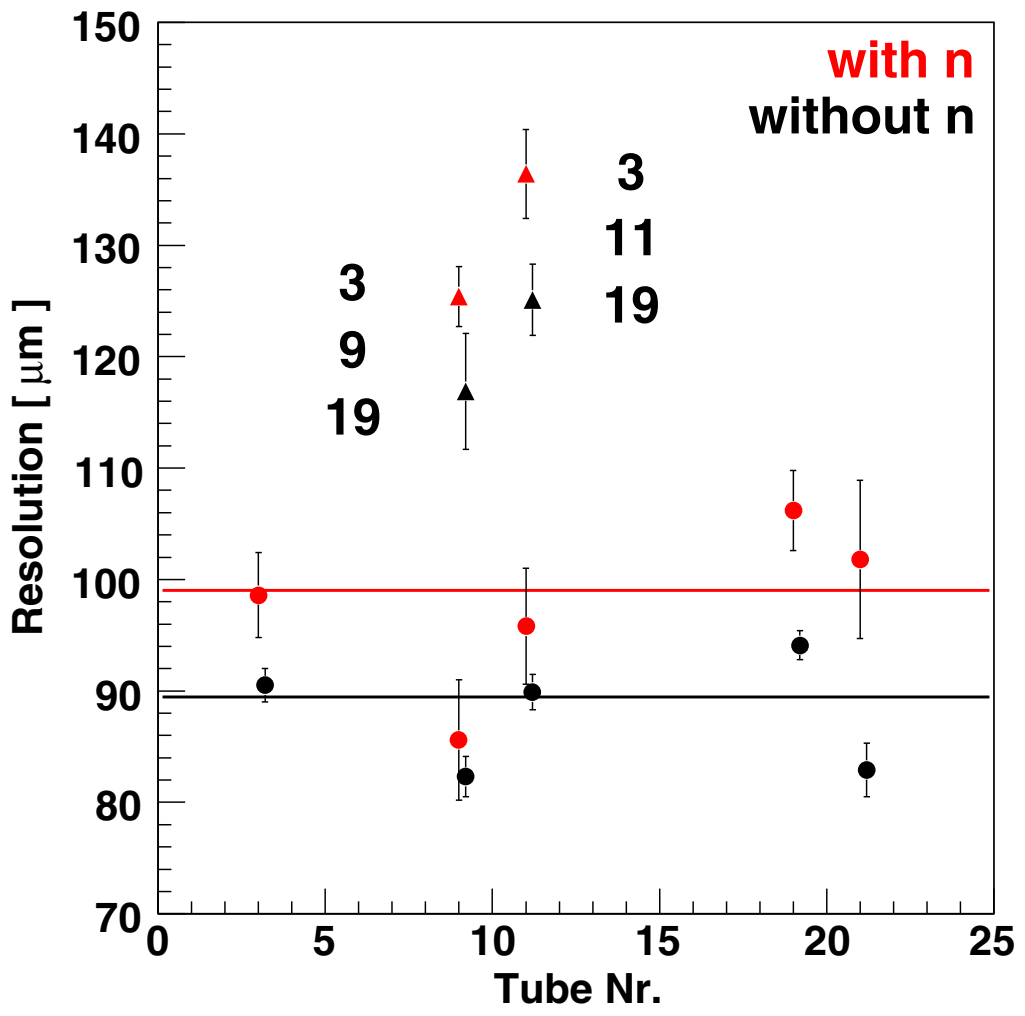

Fig. 11. Resolution of single MDT tubes without and with neutron irradiation. Two methods were used: 1. The comparison of drift-radii with the predictions of the tracks through the silicon strip detectors (full circles). 2. The comparison of 3 correlated drift-radii (triangles).

## References

[1] ATLAS Technical Design Report, ATLAS-TDR-010, CERN-LHCC-97-022, 1997
[2] S. Baranov *et al.*: ATL-Gen-2005-001, CERN (2005)
[3] M.Deile, et al., Nucl. Instr. and Meth. A518 (2004) 65
M.Deile, et al., Nucl. Instr. and Meth. A535 (2004) 212
[4] T. A. Mueller, diploma thesis, LMU, 2006
[5] A. Mlynek, diploma thesis, LMU, 2006
[6] http://geant4.web.cern.ch/geant4/
[7] M. Drosg: IAEA-NDS-87 Rev. 9, May 2005
[8] http://www.fluka.org
[9] D.E.Groom, Passage of Particles Through Matter, Eur.Phys.Journal C, Vol.3, 1-4, (1998) 144